# Token-curated registry (TCR) in a scholarly journal: blockchain meets invisible colleges[1]

**Corresponding author:**

Artyom Kosmarski, Laboratory for the Study of Blockchain in Education and Science (LIBON), State Academic University for the Humanities (GAUGN), Maronovskiy lane 26, Moscow, Russia, 119049.

https://orcid.org/0000-0001-8475-0754

E-mail: kosmarski@gaugn.ru

**Second author:**

Nikolay Gordiychuk, Laboratory for the Study of Blockchain in Education and Science (LIBON), State Academic University for the Humanities (GAUGN), Maronovskiy lane 26, Moscow, Russia, 119049.

E-mail: gordiychuk@gaugn.ru

https://orcid.org/0000-0002-2117-7601

Abstract

In this paper, we propose a novel framework for a scholarly journal – a token-curated registry (TCR). This model originates in the field of blockchain and cryptoeconomics and is essentially a decentralized system where tokens (digital currency) are used to incentivize quality curation of information. TCR is an automated way to create lists of any kind where decisions (whether to include N or not) are made through voting that brings benefit or loss to voters. In an academic journal, TCR could act as a tool to introduce community-driven decisions on papers to be published, thus encouraging more active participation of authors/reviewers in editorial policy and elaborating the idea of a journal as a club. TCR could also provide a novel solution to the problems of editorial bias and the lack of rewards/incentives for reviewers. In the paper, we discuss core principles of TCR, its technological and cultural foundations, and finally analyze the risks and challenges it could bring to scholarly publishing.

---

[1] This is a preprint (submitted version) of a paper published on 20 May 2020 in Learned Publishing. The link to final version of the paper: https://onlinelibrary.wiley.com/doi/full/10.1002/leap.1302

- There is a demand for the digital transformation of scholarly communication, new practices, and frameworks – while keeping the critical principles of journals as clubs or communities.
- We propose a model for an academic journal: a token-curated registry (TCR), based upon ideas from the field of blockchain and cryptoeconomics.
- TCR is a decentralized system where decisions are made by voting that brings benefit or loss to voters.
- TCR in a journal could act as a tool to distribute the task of reviewing to a broader community of scholars concerned about the development of a journal.
- Successful implementation of this framework could transform a journal towards becoming both an economically-minded and reputation-based self-regulatory system, balancing the interests of all stakeholders – authors, reviewers, editors, readers.

1. Introduction: academic publishing in flux

The field of scholarly communication (scholcom), and academic publishing, in particular, is passing through turbulent transformations, and even its near future appears uncertain. Arguably the critical disruptor is the breakaway from subscription model towards open access and hybrid approaches, spearheaded recently by Plan S. This initiative by major research agencies and funders from twelve EU countries obliges scientists to channel their output towards open access journals by 2021 (Else, 2019). Although the movement towards OA has a few just causes beyond it – resentment at the oligopoly of large publishers (Else, 2018) and endorsement of the principles of open science (Pulverer, 2019) – the inevitability of OA future is not beyond question. It puts a heavy burden of article processing charges (APCs) on scholars (pay to publish model). This new economic reality, with APCs ranging from €300 to €5000, would exacerbate inequalities between global North and global South (Poynder, 2019; Debat & Babini, 2019), as well as between "rich" (i.e., supported by generous grants) and poor authors (PSA Response, 2019). Still, the numbers of OA journals and OA articles continue to grow, although less spectacularly than in the early 2010s (Green, 2019, Fig. 1). Regardless of these shifts on the global scale, open access has brought about an entirely new economic model: financial decisions are taken not by a small group of librarians and administrative staff, but by a larger body of authors, who weigh quality and APCs, deciding where to send their paper. Journals, in turn, have to compete for authors, driving APCs down or inventing other incentives (this economic model is elaborated in (MacKie-Mason, 2016).

Another area of concern and change is the institution of peer review, arguably the primary source of added value in a journal article (making it different from a preprint). A longitude survey shows that the numbers of scientists arguing that peer review hampers scholcom has risen from 19% in 2007 to 26% in 2015, while the confidence in its positive role has dropped from 85% to 75% in the same period (Ware, 2008; Ware, 2016). Chief complaints about the current peer review system may be summarized as follows: lack of transparency; ample opportunities for arbitrariness; reviewers' work is invisible, insufficiently acknowledged and rewarded.

Peer review as an institution has few checks against the arbitrariness of the editor - by choosing reviewers, they may easily achieve the desired result (positive or negative reviews). Often an editor makes the final choice when the reviewers' opinions differ. Only 10% of biased editors reduce the quality of published articles by 11% (Wei Wang et al., 2016). Further, blind peer review is frequently lop-sided: it is easy for the reviewer to guess the author from the references or even from the content of the manuscript and, while maintaining anonymity, unleash their ill will. According to one study, the anonymity of the author can be preserved in less than 50% of cases (Snodgrass, 2006).

However, these concerns are dwarfed by a more pressing issue – the lack of acknowledgment and rewards for reviewers. The number of articles in scientific journals is growing every year, and the demand for reviews grows with it. Many scientists receive five or more requests to write a review every week (Fox & Petchey, 2010). A survey conducted by Taylor & Francis showed that 60% of editors struggle to find competent reviewers (Taylor & Francis, 2016). In 2015, the demand for reviewers in the biomedical sciences was 15-24% above supply; 20% of scientists wrote between 69% and 94% of reviews; 70% of scientists spent 1% or less of their time writing reviews, and 5% of scientists spent 13% or more (Kovanis, 2016). In sociology and political science, 16-18% of scientists write 50% of reviews (Saideman, 2017).

In order to mitigate these imbalances and to improve the quality of reviews and encourage authors, various ways of encouraging reviewers have been proposed. Experiments have been carried out, with regular cash payments (Chetty et al., 2014), specific scientific "currencies" – academic dollars (Pruefer, J. & Zetland, D., 2010) or PubCreds (Fox & Petchey, 2010) None of these ideas, however, has yet gained significant acceptance.

Closer to the point of this paper, amid a multitude of experimental approaches and reform agendas for scholcom, two ideas are most pertinent to the model of an academic journal outlined below. First is Toby Green's (2019) appeal for the ground-up digital transformation of scholcom – not just digitalizing current processes, but changing the whole ecosystem. Green suggests radically cutting the number of articles published (in favour of faster and cheaper preprints) and proactive behaviour on the part of editors (they would look out for the lucky few papers to be published). Second, and more important, is the idea of a scholarly journal as a club, "where a group of scholars works together to understand their domain and share common language and knowledge as markers of insider/outsider status," put forward by Hartley, Potts et al. (2019). Journals are not about technology and outputs, but about a group of scholars interested in a particular intellectual field – having a stake in it, in Bourdieusian terms (Bourdieu, P., & Wacquant, L. J. D., 1992). Scholars are keen on defining the boundaries of their community of interest, and the work of reading, writing, and reviewing the papers for a journal allows constructing such a community/club.

Hereafter we would argue that blockchain, and token-curated registries, in particular, could provide a technologically savvy framework for running an academic journal as a club, a visible manifestation of an invisible college. Community-driven decisions on papers to be published act as a device to delineate the values, positions, and boundaries of a particular journal.

2. Blockchain for science, cryptoeconomic primitives, token-curated registries: how it works

Before getting down to the particulars of a new framework for academic journals, we would give a brief overview of the blockchain (distributed ledger technology, DLT) and its current applications in scholarly publishing.

A distributed ledger is a set of data blocks connected by cryptographic tools, to make it impossible to change the content of one block without interfering with all the others. In that digital ledger, recorded transactions are transparent, and the information about them is stored across many computers. This approach to handling data (decentralized and distributed) prevents retroactive altering of data.

The attractiveness of blockchain to industry rests primarily upon its promise to make data stable, transparent, and decentralized. However, it is not the data handling but the social appeal of the blockchain (Atzori, 2015; Jun, 2018) that has attracted the attention of academia. Principal advantages of blockchain in this perspective, apart from stability and verifiability of data, are the guarantee of trust in the trustless environment and successful peer-to-peer interactions without the need for a central governing body ("the third party"). These features dovetail with the logic of modern science: it is international, decentralized, horizontal, a republic of science, in a sense (Polanyi, 1962). An analysis of blockchain applications in science, from research data to funding, is beyond the scope of this paper. For a concise review of current implementations and challenges facing them see (van Rossum, 2017; Janowicz et al., 2018).

The core principles of the blockchain (decentralization and disintermediation) naturally suggest an idea of an independent publishing platform where authors and reviewers interact directly with each other in a peer-to-peer network, without any middlemen, such as (allegedly) avaricious publishers. Not surprisingly, this idea has been so appealing that virtually every blockchain start-up in science has promised an open access platform (scienceroot.com, eurekatoken.io, pluto.network, orvium.io).

However, the success of these platforms has been limited so far, with less than a hundred papers in each: scientists prefer to publish in established journals relevant to their research communities. This trend is well-known in the sociology of innovation (e.g., Dahlin, 2014): the revolutionary benefits of new technology are insufficient to draw users away from customary practices. Probably a more promising path to introduce DLT would be to move beyond solutions for existing woes (e.g., removal of intermediaries or time-stamping a text on blockchain to establish priority and secure IP rights) towards new ideas for shaping the relations between key actors of scholcom, namely authors, reviewers, and editors.

This paper examines neither blockchain *per se* (a decentralized and distributed digital ledger) nor cryptocurrencies (digital assets, such as Bitcoin, employing cryptography to secure financial transactions), but the so-called **cryptoeconomic primitives** - self-sustaining incentive systems that allow their participants to coordinate efforts and achieve shared goals. In other words, cryptoeconomic primitives are token-based incentive systems that enable coordination to achieve a shared goal via the use of various economic and cryptographic mechanisms (Jacob, 2018).

The basic unit of these systems is a token – a unit of value programmed and operated on any blockchain. There is another definition of a token, more geared towards its commercial properties: "a unit of value that an organization creates to self-govern its business model, and empower its users to interact with its products, while facilitating the distribution and sharing of rewards and benefits to all of its stakeholders" (Mougayar, 2017). This definition, however, narrows token down to an electronic currency of sorts. In contrast, it could be framed as a digitally secure measure of anything, not just money or material value – respect, reputation, novelty, whatever a particular community employing these tokens sees fit.

One of such cryptoeconomic primitives is the token-curated registry (TCR) – a decentralized system in which tokens are used to incentivize quality curation of information presented in the form of lists. The concept of token-curated registries was first proposed in 2017 by Mike Goldin, a software engineer at ConSensys, a blockchain technology company.

**In brief, TCR is an automated way to create lists of any kind where decisions (whether to include N on the list) are made through voting that brings benefit or loss to voters.** The advantage of TCR is based on the following idea: the free market and material interest of individuals provides better information than non-transparent centralized systems. The system encourages the search for and entry of reliable information into a list and disincentivizes the entry of unreliable information. "So long as there are parties which would desire to be curated into a given list, a market can exist in which the incentives of rational, self-interested token holders are aligned towards curating a list of high quality. Token-curated registries are decentrally-curated lists with intrinsic economic incentives for token holders to curate the list's contents judiciously" (Goldin, 2017).

The lists in the TCR system may be made up of everything – universities, restaurants, movie stars, websites, videos, etc. There are three user types in TCR, and each type has its interests and incentives.

1. Candidates want to be included in the list (e.g., "Best Chinese restaurants in Leicester");

2. Consumers are lay users who need high-quality information from a list (e.g., those choosing a fancy Chinese restaurant to go out);

3. Curators or token holders are the voters who can earn something from their honest curation (whether to accept a specific restaurant in the list or reject it).

To become a curator, one needs to purchase an intrinsic token and thus acquire the right to vote in the TCR system (alternatively, the tokens may be allocated for free to recognized experts in the field). To become a candidate, one has to purchase tokens and make a $D_{cand}$ deposit. After a candidate's tokens are deposed, the waiting period starts (Fig. 1). During this period, the application may be challenged by a curator. In this case, they are obliged to stake a $D_{chall}$ counter deposit. If the application goes unchallenged, the candidate is included in the list. Otherwise, the voting begins: curators may vote for or against the inclusion of a candidate on the list. The voting process is limited in time (e.g., one week). If a majority votes for (i.e., the challenger loses, and the candidate wins), the challenger forfeits her deposit, which is distributed equally among those who voted for the inclusion. If a majority has voted against the candidate, her deposit is also forfeited and distributed among those curators who had opposed the candidate.

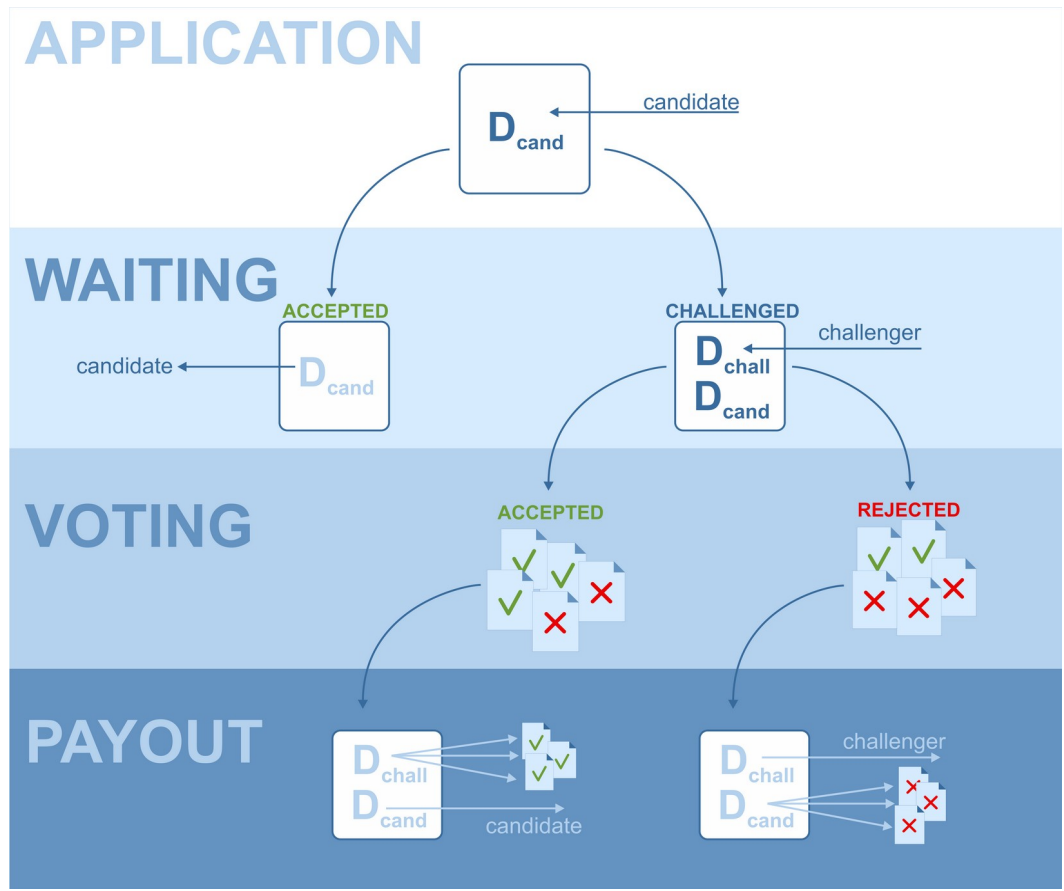

*Figure 1. Voting process within a TCR framework. Image credit: Natalia Deriugina.*

Ideally, the system maintains a precarious balance. If curators accept or turn down candidates whimsically, consumers would find the list inaccurate and cease using it. Then candidates would lose interest in applying, and the value of a token (proportionate to the number of aspiring candidates who stake a deposit) would drop, thus hurting the interests of curators. Also, constant rejection of candidates would benefit curators in the short term (they share the deposits). Still, in the long run, candidates would become reluctant to apply, and the token value would again drop. So, "rather than relying on a centralized entity or small group to curate a list, TCRs create a distributed network of participants driven by an effective framework of incentives. Benefits range from removal of bias and bribery potential of the list curators to the ability to have a constantly updating representation of an important subset of data within a specific market or an amalgamation of related metrics" (Curran, 2018).

This protocol is just a generic model for a TCR. Many modifications have been proposed and implemented so far. A list might become more complex, with layers and rankings for candidates – ordered TCRs (Pereira, 2018), upvotes/downvotes (Gajek, 2018), tournament-like layered TCRs (McConaghy, 2018b). Curators could vote on objective conditions rather than on candidates – e.g., not on inclusion of college N into the list of top 20 US colleges, but on the relative importance of a graduate's salary vis-à-vis average H-index of professors (Clark, 2018). Of course, weak points of TCR technology (e.g., trolling by knowingly proposing foul candidates, or coin flipping – voting both for and against instead of making sober assessments of the issues at stake), and ways to overcome them have been discussed as well (Gerbrandy, 2018; Balasanov, 2018).

Crucially, TCR as a social and economic model rests upon a set of assumptions, common to many cryptoeconomic endeavours:

1) Incentive design: the idea that one can make people do the right things if incentives within a tokenized system are aligned right enough. "The blockchain community understands that blockchains can help align incentives among a tribe of token holders… But the benefit is actually more general than simply aligning incentives: you can design incentives of your choosing, by giving them block rewards. Put another way: you can get people to do stuff, by rewarding them with tokens. Blockchains are incentive machines" (McConaghy, 2018a)

2) Skin in the game – a person's decision gets more sound if they have something to win or lose from it. Skin in the game distinguishes TCR from conventional voting or commenting online. This concept has been recently popularized by Nassim Taleb (Taleb, 2018).

3) The related idea of private self-interest (curators' desire to maximize their profits) turning into a public benefit (lists with high-quality information), It has strong roots in classic economic liberalism, e.g., (Horne, 1981).

4) Wisdom of the crowd, a concept implying that the collective opinion of a group of individuals gives more accurate information than the opinion of a single expert (pioneered in institutions such as trial by jury, Wikipedia, or Quora). For applications in blockchain, see (Muehlemann, 2018).

3. TCR-based scholarly journal: participatory governance and skin in the game

Next, we will describe how TCR technology may be employed in an open access journal, what pressing issues of scholarly publishing it could resolve, and which new challenges it would pose. In a nutshell, TCR candidates are the authors eager to see their papers accepted. Curators are members of an "invisible college," or a club (Hartley, Potts, et al., 2019) behind a particular journal, anxious to uphold its reputation and keep up its development. Finally, consumers are readers.

Before submitting a paper, an author purchases a sum of tokens for their fiat currency (for example, the equivalent of €300), depositing them as APC. When a manuscript is uploaded to the online platform, reviewers (curators) may vote whether to accept or to reject it.

A curator cannot take part in the voting without first submitting a review – a short or a longer one, or an evaluation of qualities of the paper on a 10-point scale, as the editorial board sees fit. Subsequently, the reviews would be open not only to the authors, but also to other curators, and, if a journal opts for an open peer review model, to the general public. Making a review (even a short one) a condition for voting is meant to deter irresponsible and haphazard voting on the part of the curators – this rule could be taken down, of course.

Further on, two options are possible. In the first case, curators/reviewers have an opportunity to study all the reviews done before, and to evaluate the arguments of their colleagues in favour of or against the publication. Otherwise, curators have to vote simultaneously with submitting their reviews.

If the number of votes "for" is not equal to zero, and is identical to or exceeds the number of votes "against," an article is accepted for publication (a journal may also introduce a minimal amount of votes required for a paper to be accepted). Curators who have voted in favour are rewarded with tokens from the deposit (in equal stakes). If the vote rejects an article, the tokens are distributed among the curators who voted against it. In each case, a certain portion of the deposit is used to cover the expenses of the journal (editing, overhead costs, etc.).

In this way, a system of incentives is embedded into the workings of an academic journal, enhancing its qualities of a self-regulatory system. Voters (curators) are interested in getting the best articles, as it would boost their reputation as the community behind that journal. They are also encouraged to participate more actively and responsibly in the review process, as this activity would reap tangible benefits – they have their skin in the game! If curators are too complacent and accept too many articles, including weak ones, the reputation of a journal would suffer, and new authors would be less willing to publish there. On the contrary, if curators set too high a bar on acceptance, authors would be reluctant to submit their papers (the risk of losing their deposit and not having an article published would be too serious).

There are other important features embedded in this model of a TCR-based scholarly journal. First of all, every interested and competent member of the professional community has an opportunity to write a review. In a conventional peer-reviewed journal, the editorial board chooses two or more reviewers. In contrast, we propose registering a new manuscript in a closed database, where it becomes available to all potential reviewers. Afterward, community members can choose the papers most relevant to them, for voting and reviewing. This framework might accelerate the process of peer review, ease the burden of editors, and raise the transparency of a journal's workflow (e.g., by curtailing editors' bias in choosing reviewers). It also dovetails with the recent trend of the so-called collaborative peer review (as practiced in Cell Press journals, for instance) Finally, it would partially solve the issue of overburdened reviewers lacking recognition and remuneration for their labours (see above, section 1).

Next, this model implies that a journal is run by an active community, or club, to use the concept by Hartley, Potts, et al. (2019). This, of course, raises the question of what constitutes such a community. One idea would be to grant voting/curation/reviewing rights to some or all the authors who had published their papers in a journal before. The fact of getting one's paper published indicates that one understands the journal's research field, aims, and scope. Therefore, such a person would act as a sensible reviewer.

Our approach allows authors to play a more active role in how "their" journal is getting on: they would influence the editorial policy more actively. A continually expanding community of scholars coalesces around a journal, investing their time and effort in its development. This model is inspired by the principles of participatory governance (participatory democracy) – a concept that emphasizes citizen participation in deliberating, making, and monitoring decisions on the issues somehow relevant to their well-being. This concept was pioneered in the 1970s (Pateman, 1970) but has recently experienced a rebirth, as the Internet and new digital technologies have made it easier to "embed" participatory governance in everyday life. (Poletta, 2016). The authors, having contributed to the build-up of a journal's reputation, become the stakeholders interested in its well-being. TCR framework implies that curators' benefit will suffer if a journal accepts low-quality papers. Therefore, as "owners" of the journal's reputation, they should be able to participate directly in the editorial policy and be responsible for its results.

Of course, running a journal on TCR entails a few risks. A curator, for instance, might write haphazard reviews and vote at random, assuming that such "coin-flipping" behaviour would be profitable enough and save them from spending time on sincere evaluation of manuscripts. The following measures may be employed to counterbalance such actions:

1) Introduction of an open review system, where curators' opinions and identities are disclosed. These texts would serve as an indicator of a reviewer's reputation.

2) Running the list of reviewers/curators also as a TCR. In this case, the community might vote to expel the curators whose behaviour is deemed unethical or otherwise detrimental to the values of a given journal.

3) Establishing a limit for the number of papers any curator might review (in a month or six months, for instance). This limitation would encourage more responsible reviewing and prevent the formation of a clique of "professional" curators who are only concerned with voting for/against papers for their short-term profits, regardless of the interests of the wider community.

Another source of friction would probably be the formation of cliques among curators, strong enough to form a voting block necessary to accept or block any paper. Such formation of interest groups (as well as the voter apathy) is arguably the inevitable outcome of introducing democratic politics into any sphere, not only scholcom. In our opinion, a community around a journal could imagine and implement a set of rules and incentives to counterbalance such side effects, as it is done in any democratically governed institution.

4. Conclusion

Introducing TCR algorithms into a scholarly journal would be an interesting experiment. It could partially ease the burden on reviewers (by distributing these labours across a wider community and by incentivizing the curators). It also holds promise to make a journal a self-regulatory system, adjusting the interests of all participants – authors, reviewers, and readers. Finally, the TCR model offers a tool to engage a broader community behind a particular journal, make them active citizens working to define editorial policy.

Precisely this feature enables TCR to reinvigorate the journal as a club. "The shared good in production is mutual attention to an idea, which is the good allegedly supplied by the publishing intermediary in the idealized form of generalized attention. What scholars are 'purchasing' is shared access to the benefits of other smart, like-minded scholars who are implicitly part of an open team production exercise… Club membership is by itself an incentive to improvement, of each and of all… For instance, authors and other members care a great deal about who else is a member, who else is an author, and exactly how the attention of specific readers is apportioned… Having the right people publish in the journal adds prestige benefits for all members" (Hartley, Potts, et al., 2019). In this sense, TCR is the tool to increase that public good; to provide a flexible instrument to filter the "right people" and to define the profile and reputation of a journal, thus acting as the incentive to improvement.

However, with community building as the fundamental goal of the TCR framework, its other face is the economic reasoning (rewards for curation, skin in the game, monetary gain as an indicator of self-alignment with the values of the community). To what degree these two goals might be aligned, and whether financial incentives would serve the reputation economy (Green, 2019) of a journal rather than spoiling it, is an open question. We hold, however, that the TCR model is worthy of experimenting. Implementing it in real-life conditions, even on a small scale, would bring to light its strengths and weaknesses, as well as provide valuable insights on the interplay between the stakeholders of a scholarly journal.

Disclosure statement

No potential conflict of interest was reported by the authors.

van Rossum, J. (2017) Blockchain for Research. Perspectives on a New Paradigm for Scholarly Communication. Digital Science Report. Retrieved from: https://doi.org/10.6084/m9.figshare.5607778

Saideman, S. (2017, November 13). Are Potential Peer Reviewers Overwhelmed Altruists or Free-Riders? New data reveal great inequality in peer reviewing in the social sciences. *Duck of Minerva*. Retrieved from: http://duckofminerva.com/2017/11/are-potential-peer-reviewers-overwhelmedaltruists-or-free-riders-new-data-reveal-great-inequality-in-peer-reviewing-in-thesocial-sciences.html

Snodgrass, R. (2006). Single-versus double-blind reviewing: an analysis of the literature. *ACM SIGMOD Record, 35(3)*, 8-21. https://doi.org/10.1145/1168092.1168094

Taleb, N. N. (2018). *Skin in the Game: Hidden Asymmetries in Daily Life*. New York: Random House.

Taylor & Francis Group. Peer review in 2015: a global view. A white paper. Retrieved from: http://authorservices.taylorandfrancis.com/peer-review-global-view/

Ware, M. (2008). *Peer review: benefits, perceptions and alternatives. Summary report*. Publishing Research Consortium; 2008. Retrieved from http://publishingresearchconsortium.com/index.php/prc-documents/prc-research-projects/35-prc-summary-4-ware-final-1

Ware, M. (2016). *Peer Review Survey 2015*. Publishing Research Consortium. Retrieved from http://publishingresearchconsortium.com/index.php/134-news-main-menu/prc-peer-review-survey-2015-key-findings/172-peer-review-survey-2015-key-findings

Wei Wang et al. (2016) Editorial behaviors in peer review. *SpringerPlus, 5(1)*, 903. https://doi.org/10.1186/s40064-016-2601-y
17